\documentclass[traditabstract,letter]{aa}

\usepackage{graphicx}
\usepackage{txfonts}
\usepackage{longtable}
\usepackage[labelfont=bf]{caption}
\usepackage{lscape,supertabular}
\usepackage{natbib}
\usepackage{hyperref}
\usepackage{subfig}
\usepackage{enumitem}
\usepackage{inputenc}
\usepackage[range-units = brackets,tophrase={-},seperr]{siunitx}
\begin{document}

   \title{The long X-ray tail in Zwicky 8338}

   \author{G. Schellenberger\thanks{gerrit@uni-bonn.de; Member of the International Max Planck Research School (IMPRS) for Astronomy and Astrophysics at the Universities of Bonn and Cologne}
          \and
          T. H. Reiprich
	  }

   \institute{Argelander-Institut f\"ur Astronomie, Universit\"at Bonn, Auf dem
               H\"ugel 71, 53121 Bonn, Germany
             }

  \abstract
   {The interaction processes in galaxy clusters between the hot ionized gas (ICM) and the member galaxies are of crucial importance in order to understand the dynamics in galaxy clusters, the chemical enrichment processes and the validity of their hydrostatic mass estimates. Recently, several X-ray tails associated to gas which was partly stripped of galaxies have been discovered. Here we report on the X-ray tail in the $3\,$keV galaxy cluster Zwicky 8338, which might be the longest ever observed. We derive the properties of the galaxy cluster environment and give hints on the substructure present in this X-ray tail, which is very likely associated to the galaxy CGCG254-021. The X-ray tail is extraordinarily luminous ($\SI{2e42}{erg\,s^{-1}}$), the thermal emission has a temperature of $\SI{0.8}{keV}$ and the X-ray luminous gas might be stripped off completely from the galaxy. 
   	From the assumptions on the 3D geometry we estimate the gas mass fraction ($< 0.1\%$) and conclude that the gas has been compressed and/or heated.}
\titlerunning{The X-ray tail in Z8338}
\authorrunning{Schellenberger \& Reiprich}
   \maketitle
\section{Introduction}\label{sec:intro}
Galaxy clusters are important tools for cosmology. They consist of galaxies, the intra cluster medium (ICM), which is a very hot plasma (several $\SI{e7}{K}$), and Dark Matter. This hot gas is emitting mainly line and free-free emission in X-rays. 
Although the ICM undergoes many processes that affect the composition and thermal structure, a simplistic assumption is to treat galaxy clusters as relaxed objects.
Understanding all these processes, which play a key role in the evolution, is essential to get robust constraints on parameters like the gas and total mass. One important process to look at, is the in-fall and dynamics of member galaxies as initially described by \cite{1972ApJ...176....1G}. Halo gas and the cold inter stellar medium (ISM) from the galaxy can then be stripped off and interact with the ICM. Part of the gas from the galaxy is then used for new stars, either in the galactic halo or outside the galaxy in the ICM (see e.g., \citealp{2007ApJ...671..190S}). Simulations of the interaction between ISM and ICM (e.g., \citealp{1999MNRAS.310..663S,2009A&A...499...87K,2012A&A...544A..54S,2015ApJ...806..103R}) predict a leading bow shock and tail behind the galaxy, which are both visible in X-rays. The highest chance to find these structures is in cool nearby clusters with blue galaxies. Since not always an X-ray counterpart is found when e.g., an HI tail is detected (see \citealp{2005A&A...437L..19O}), the process of stripping off gas from the host galaxy is still not understood completely and should be analyzed in more detail. Unfortunately only very few detections in X-rays have been made so far (e.g., \citealp{2004ApJ...611..821W,2005ApJ...621..718S,2005ApJ...630..280M,2006ApJ...637L..81S,2008ApJ...688..208R,2008ApJ...688..931K}).

In our short Chandra observation of the galaxy cluster Z8338, we identify a member galaxy exhibiting one of the longest X-ray tails known to date (at least 76 kpc length). It turns out that this object is already listed in the ROSAT catalog (\citealp{1999A&A...349..389V}) with the Name \textit{1RXSJ181030.0+495615}. The projected distance of this galaxy from the main cluster center is $\SI{310}{kpc}$. Surprisingly, the galaxy has probably lost all of its X-ray emitting gas very recently -- the peak of the emission is $\SI{40}{kpc}$ offset from the galaxy center, this has never been seen before so clearly. 
Moreover, we see hints for a bow shock by an increased temperature in a region in front of the X-ray tail.
Here, we present the parameters like the thermodynamic structure derived from the short observation of this very long and bright tail. Likely it is the longest X-ray tail with the largest separation from the host galaxy ever detected. We assume a $\Lambda$CDM cosmology with $\Omega_{\rm m} = 0.3$, $\Omega_\Lambda = 0.7$ and $H_0 = \SI{71}{km\,s^{-1}\,Mpc^{-1}}$, to be consistent with \citealp{2011A&A...534A.109P}(P11). All uncertainties are at the 68\% confidence level.

\section{Data analysis and results}
For the following analysis the CIAO Software package 4.7 and CALDB 4.6.7 as well as the HEASOFT tools 6.17 including Xspec 12.9 (\citealp{arnaud_xspec}) were used. Since we are dealing with regions of few counts, we use the Xspec implemented C-statistics (\citealp{1979ApJ...228..939C}). For each spectral fit we verified our parameter estimates and degeneracies by performing an MCMC within Xspec, so all derived quantities like the luminosity were calculated using the distributions of the source parameters (temperature, abundance, normalization and in some cases redshift).
The steps for the data reduction follow the same structure as described in \cite{Schellenberger2015}.
The solar abundances are set to the values given by \cite{2009ARA&A..47..481A}. The influence of the abundance table, especially at lower plasma temperatures on the best fit parameters, are shown e.g., in \cite{Lovisari2015}.

\begin{figure*}[t]
	\centering
	\includegraphics[trim=50px 250px 50px 290px,clip,width=0.77\textwidth]{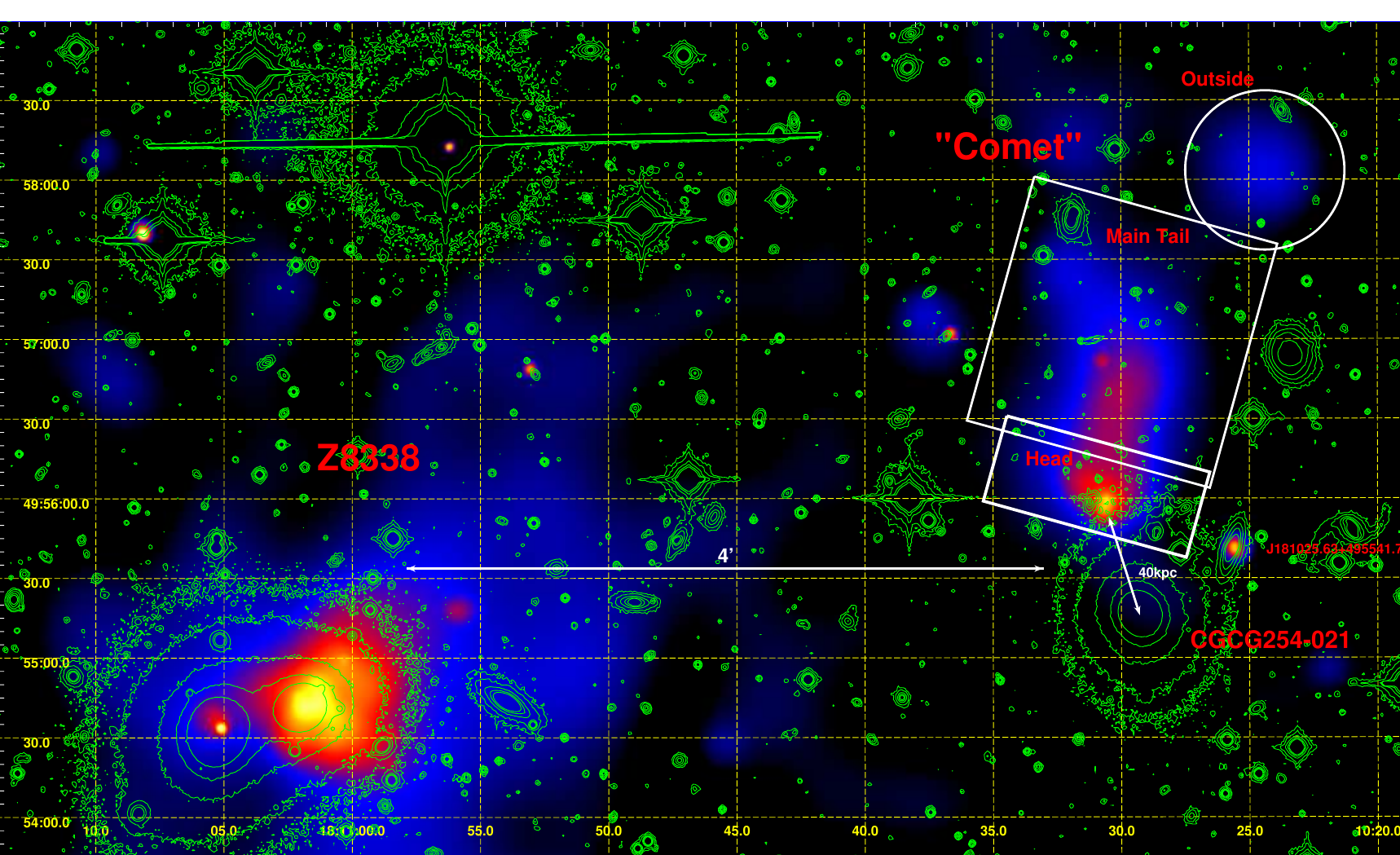}
	\caption{\small Adaptively smoothed X-ray image with V-band contours from WINGS (\citealp{2006A&A...445..805F}) in green. The distance of the Comet from the cluster center (Z8338) is $\SI{5.3}{\arcmin}$ ($\SI{310}{kpc}$). The tail of the Comet has roughly a length of $\SI{1.3}{\arcmin}$ ($\SI{76}{kpc}$). Two close member galaxies (CGCG254-021 and J181025.63) are marked as well, we assume that the comet-like X-ray emission is due to gas stripped from CGCG254-021.}
	\label{fig:smooth}
\end{figure*}

\subsection{The Cluster}
The galaxy cluster of interest for this work is listed in P11 with the following basic properties:
\begin{itemize}
	\item The name is given as ZwCl8338 or MCXC J1811.0+4954.
	\item The equatorial coordinates are \\RA = $\SI{272.7504}{\degree}$, DEC = $\SI{49.9111}{\degree}$ (J2000).
	\item $L_{500} = \SI{5.3e43}{erg\,s^{-1}}$ in the $0.1-2.4\,{\rm keV}$ energy band, which corresponds to $\SI{3.4e43}{erg\,s^{-1}}$ in $0.5-2{\rm keV}$ band.
	\item $M_{500} = \SI{1.3e14}{M_\odot}$.
	\item $R_{500} = \SI{0.767}{Mpc}$.
	\item Redshift $z = 0.0501$ (as given in \citealp{2000ApJS..129..435B}).
\end{itemize}
While analyzing our $\SI{8}{ks}$ Chandra observation from Observation Cycle 14 (OBSID 15163) pointed at this cluster we discovered a comet-like structure to the west (Fig. \ref{fig:smooth}). 

Analyzing the cluster Z8338 itself by extracting spectra reveals some details on the ICM environment. In the following, the ICM  emission is described by an \verb|apec|-model (AtomDB 2.0.2) combined with a \verb|phabs|-model to account for the Galactic absorption ($N_{\rm H} = \SI{4.8e20}{cm^{-2}}$, \citealp{2013MNRAS.tmp..859W}). Several regions around the cluster center (excluding point sources and the Comet) are fitted (see Table \ref{tab:cluster}).

\begin{table}[tb]
	\centering
	\footnotesize
	\begin{tabular}{cccc}
		\hline
		Region & k$T$ & $Z$ & $L_{0.5-2}$\\
		       & [keV] & $Z_\odot$ & $\SI{e43}{erg\,s^{-1}}$\\
		\hline
		\rule{0px}{9px}$\SIrange{0}{0.5}{\arcmin}$ & $1.89^{+0.09}_{-0.10}$	& $0.94^{+0.34}_{-0.25}$	& $0.40^{+0.02}_{-0.02}$\\
		\rule{0px}{9px}$\SIrange{0}{4}{\arcmin}$ & $3.09^{+0.16}_{-0.15}$	& $0.64^{+0.17}_{-0.14}$	& $1.48^{+0.03}_{-0.04}$\\
		\rule{0px}{9px}$\SIrange{4}{7}{\arcmin}$ & $3.61^{+0.42}_{-0.37}$	& $0.35^{+0.26}_{-0.19}$	& $0.84^{+0.03}_{-0.04}$\\	
		\rule{0px}{9px}$\SIrange{0}{10}{\arcmin}$ & $2.98^{+0.17}_{-0.16}$	& $0.43^{+0.13}_{-0.11}$	& $2.84^{+0.06}_{-0.05}$\\		
		\hline
	\end{tabular}
	\caption{\small Temperature, abundance of heavy elements and luminosity for the cluster in different regions using a redshift of 0.05.}
	\label{tab:cluster}
\end{table}

Leaving the redshift free to vary in a region with high signal to noise we constrain $z = 0.060^{+0.011}_{-0.022}$, which is consistent with P11. Also the calculated luminosity within 77\% $R_{500}$ deviates only by 17\% from the value in P11. Note that for the calculation of our luminosities point sources and the Comet structure have been excluded, while this is not the case in P11.
The core region of the cluster shows a significant drop in temperature.
We find a cluster temperature in the annulus that comprises the Comet of $\SI{3.6(4)}{keV}$. Compared with the inner and outer regions of the cluster this seems to be significantly higher. 
If one splits this annulus into one east and one west sector, the latter one including the region around the Comet, we detect consistent temperatures and abundances in these two sectors.

\subsection{The Comet}
\begin{figure}[t]
	\centering
	\includegraphics[width=0.45\textwidth]{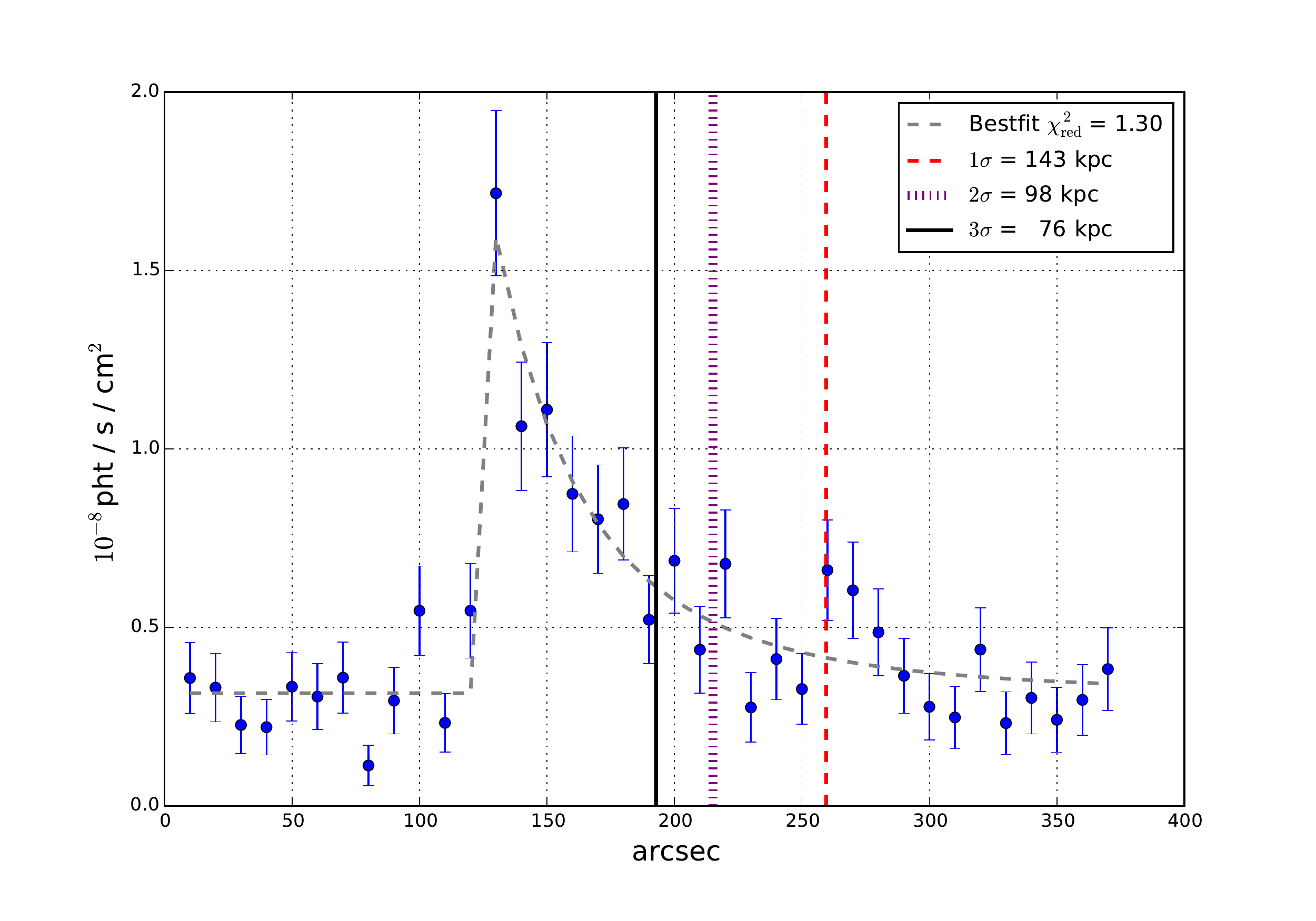}
	\caption{\small Photon flux profile across the Comet structure from south to north. }
	\label{fig:profile}
\end{figure}

The structure to the west (Comet) consists of a brighter spot to the south (``Head'') and an elongated structure of diffuse emission  (``Main Tail''). 
Estimating the length of the structure by eye from the smoothed image shown in Fig. \ref{fig:smooth}, we conclude a size of $\SI{1.2}{\arcmin}$, which corresponds to $\SI{70.5}{kpc}$ using the cluster redshift of 0.05. By extracting a profile (from equal sized boxes of $\SI{10}{\arcsec}$ width) along the tail from the (unsmoothed) exposure corrected counts image (Fig. \ref{fig:profile}), we are able to fit the function
\begin{equation}
	F(r) =
	\left\lbrace\begin{array}{r}
	c+a\cdot r^{-b}~\mathrm{, for} ~r > r_0\\
	c			  ~\mathrm{, for} ~r < r_0\\
	\end{array}\right.
\end{equation}
to the photon flux, where $a$, $b$, $c$ and $r_0$ are free parameters. Using the same redshift, we find a length of 143, 98 or 76 $\si{kpc}$, when the function reaches at $c$ plus $1\sigma$, $2\sigma$ or $3\sigma$, respectively. The flux increase in Fig. \ref{fig:profile} at around $\SIrange{260}{280}{arcsec}$ (the red dashed line) corresponds to the light peak labelled ``Outside'' in Fig. \ref{fig:smooth}. This could imply that the Comet consists of substructure and its total length extends beyond $\SI{150}{kpc}$. In this region outside the main tail (Outside in Fig. \ref{fig:smooth}) the temperature seems to be significantly higher than any part of the tail (Table \ref{tab:fits}), but still much lower than in the surrounding cluster region.
	
Unfortunately this structure has only around 300 source counts
in this observation, so it is hard to derive any detailed properties. Still we are able to obtain rough estimates for the properties of the Comet, given in Table \ref{tab:fits}. Due to the lower temperature, we decided to perform the spectral fits related to the Comet in the $\SIrange{0.5}{3}{keV}$ band. It turns out that the head is (with low significance) cooler than the tail, possibly due to a dense cool core. Overall, when accounting for the projected cluster emission, the Comet exhibits a temperature $\SI{0.77(8)}{keV}$ and a luminosity in the $\SIrange{0.5}{2}{keV}$ band of $\SI{2.0(2)e42}{erg\,s^{-1}}$. These properties are consistent with the expectations for X-ray tails (e.g., \citealp{2006ApJ...637L..81S}), only that the luminosity is almost one order of magnitude higher.
Despite the large uncertainties, there might be indications that the tail has a much lower abundance of heavy elements than the head of the Comet, which is again consistent with the head being the remnant of a cool core. 

We tried to determine the redshift from the spectrum with the highest count number and got a value ($0.062^{+0.028}_{-0.047}$) consistent with the one from the cluster spectrum, so we assume that the Comet is interacting with the ICM of Z8338. Also comparing the measured X-ray flux of the Comet with those of small galaxy groups by using the L-T relation from \cite{Lovisari2015}, we conclude a redshift range $z = 0.036^{+0.021}_{-0.012}$, which also excludes this structure to be a background cluster.

Having the gas properties like X-ray luminosity and temperature as well as the redshift of the Comet in hand, one can calculate the gas mass assuming a 3D shape. Since the information on the properties have large uncertainties and to just get a rough estimate, we assume a cylindrical shape with $\SI{1.2}{\arcmin}$ height and $\SI{0.15}{\arcmin}$ radius. We calculate $M^{\rm Comet}_{\rm gas} = \SI{1e10}{M_\odot}$.
Assuming the Comet would be a small galaxy group, one can use a scaling relation to convert the X-ray luminosity into the total mass at the structure the Comet used to reside in, assuming it has been stripped. Using the bias corrected $L-M$ relation for galaxy groups from \cite{Lovisari2015}, we estimate a total mass for the Comet structure of $M^{\rm Comet}_{\rm tot,LM} = \SI{2e13}{M_\odot}$ and a gas mass fraction of $0.04\%$. Using the $M-T$ relation from the same reference, we estimate $M^{\rm Comet}_{\rm tot,MT} = \SI{1.6e13}{M_\odot}$ and a gas mass fraction of $0.06\%$. For a total mass in this range, a gas mass fraction of $3-7\%$ is expected (\citealp{Lovisari2015}). Even when we dramatically increase our rough estimates for the cylinder (radius $\times 2$, height $\times 1.5$), we still get an upper limit for the gas mass fraction of $\sim0.1\%$. 
These values indicate that this object is far too luminous and hot than what is expected for its size. 
Reasons could be that the gas is compressed, e.g., by tidal ram pressure stripping processes, or heated by shocks or compression. However, assuming that the gas originally was attached to a small galaxy group and now almost completely left the gravitational potential, it is expected that at least the outer parts of it undergo adiabatic expansion. This would not only cool down the gas but also lower its density, making it more difficult to detect X-rays resulting in both an underestimated gas mass and underestimated luminosity and total mass. 

Using the spectrum of our observation of the cluster in the $\SIrange{4}{7}{\arcmin}$ annulus we can roughly estimate the pressure, $\SI{5.4(6)e-12}{erg\,cm^{-3}}$, and density, $\SI{8.0(3)e-28}{g\,cm^{-3}}$, around the Comet. 

Simulations in \cite{2009A&A...499...87K} show that for a relative velocity between the surrounding gas and the galaxy of $v_\mathrm{rel} = \SI{1000}{km\,s^{-1}}$ and an 8 times higher pressure and 6 times higher surrounding density, the scenario of a completely stripped gaseous disc can be explained. This might be a hint that the relative velocity of the galaxy is significantly larger. 
Due to the very low number of counts we find an interesting but only weak hint from a spectral analysis that there might be a high temperature region in front of the Comet.
A test to mirror the same region on the other side of the cluster strengthens this hypothesis since there we measure a temperature consistent with the one in this cluster annulus (Table \ref{tab:cluster}). Unfortunately we are not able to detect a surface brightness enhancement connected to the hypothesized bow shock region, which in turn weakens this idea.

\begin{table}[tb]
	\centering
	\footnotesize
	\begin{tabular}{cccccc}
		\hline
		Region & k$T$ & $Z$ & $L_{0.5-2}$ & S/N\\
		& keV & $Z_\odot$ & $\SI{e43}{erg\,s^{-1}}$&\\
		\hline
		\rule{0px}{9px}Head & $0.72^{+0.09}_{-0.09}$	& $0.25^{+0.22}_{-0.12}$	& $0.13^{+0.02}_{-0.02}$&\\
		\rule{0px}{9px}Tail & $0.93^{+0.16}_{-0.13}$	& $0.05^{+0.05}_{-0.03}$	& $0.17^{+0.02}_{-0.02}$&\\
		\rule{0px}{9px}Full Comet & $0.81^{+0.08}_{-0.07}$	& $0.09^{+0.04}_{-0.03}$	& $0.28^{+0.02}_{-0.02}$&\\
		\hline
		\rule{0px}{9px}Head* & $0.68^{+0.11}_{-0.17}$	& $0.32^{+0.93}_{-0.24}$	& $0.10^{+0.02}_{-0.01}$& 10\\
		\rule{0px}{9px}Main Tail* & $0.80^{+0.12}_{-0.11}$	& $0.06^{+0.04}_{-0.03}$	& $0.13^{+0.02}_{-0.01}$& 11\\
		\rule{0px}{9px}Full Comet* & $0.77^{+0.08}_{-0.07}$	& $0.14^{+0.08}_{-0.06}$	& $0.20^{+0.02}_{-0.01}$& 14\\
		\hline
		\rule{0px}{9px}Outside* & $1.17^{+0.14}_{-0.18}$ & 0.05 \scalebox{0.65}{fixed} & $0.04^{+0.02}_{-0.01}$ & 5 \\
		\hline
	\end{tabular}
	\caption{\small Best fit temperature, relative abundance of heavy elements, X-ray luminosity and signal to noise ratio using a redshift of 0.0501. * marks fits, where the Cluster emission in $\SIrange{4}{7}{\arcmin}$ was simultaneously fitted and accounted (only in these cases the S/N could be calculated properly since the cluster emission is accounted for in the noise).}
	\label{tab:fits}
\end{table}

\section{Counterparts in other wavelengths}

\begin{figure}[t]
	\centering
    \includegraphics[trim=20px 30px 30px 50px,clip,height=151px]{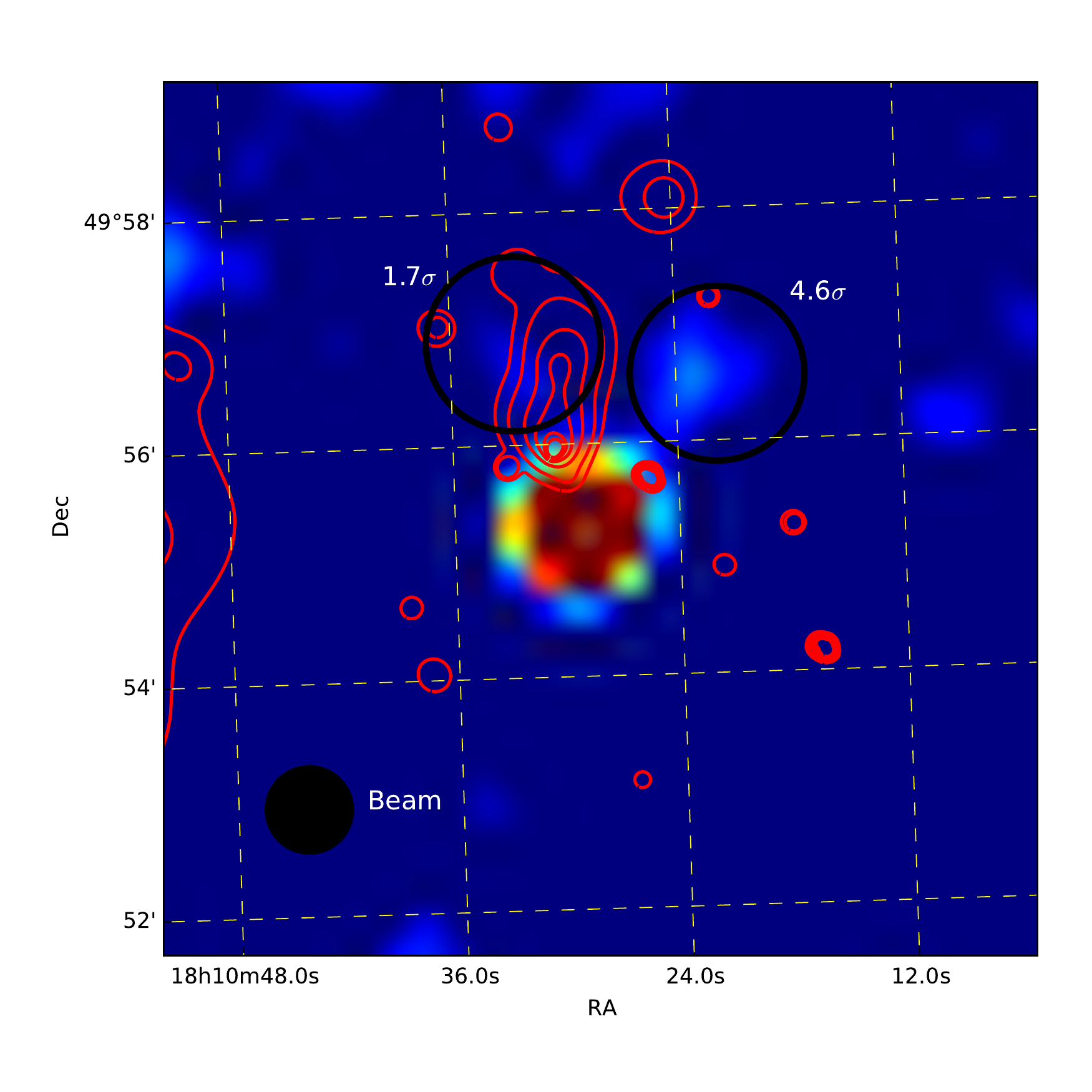} 
	\caption{\small
		$\SI{21}{cm}$-Radio continuum from NVSS showing two possible bent radio lobes qualitatively following the expected direction of motion from the X-ray morphology (red). The significance of the possible detections are indicated.}
	\label{fig:optical}
\end{figure}

Using publicly available image data from the Wide-field Infrared Survey Explorer (WISE, \citealp{2010AJ....140.1868W}) and the WINGS survey (\citealp{2006A&A...445..805F,2009A&A...497..667V,2009A&A...501..851V}) we find galaxies in the vicinity of this structure: To the south we can identify the $\SI{14.126}{mag_V}$ galaxy CGCG254-021 (also called J181029.20+495517.0, \citealp{2011A&A...526A..45F}, F11) with a stellar mass of $\SI{1.36(1)e12}{M_\odot}$, which seems unusually high. This galaxy is listed as a member galaxy of Z8338 (\citealp{2004AJ....128.1558S}; and a spectroscopic redshift of 0.0511 in \cite{2009A&A...495..707C}, in agreement with the cluster redshift in P11). According to F11, this is the galaxy with the highest mass, by far the highest star formation (SF) rate in the past ($\SI{5.5}{M_\odot \per yr}$, while no SF is detected in current age) and highest age within Z8338.
So this galaxy was a large starburst galaxy and is now clearly dominant over other objects in Z8338 in the region around the Comet.

With the colors $(\mathrm{B-V}) = 1.1$ and $(\mathrm{U-B}) = 0.95$ (\citealp{2009A&A...497..667V,2014A&A...561A.111O}) the galaxy CGCG254 is clearly an early type galaxy (\citealp{1961ApJS....5..233D}). 
Since the bolometric X-ray luminosity compared to the B-band luminosity seems to agree with what is expected for BCGs, as shown by \cite{2001MNRAS.328..461O}, one can raise the assumption that the galaxy was the BCG of a galaxy group. 
In \cite{2008ApJ...679.1162J} the authors have shown that also the X-ray and near infrared (K-band) luminosity of early type galaxies in clusters and groups are correlated. 
Applying their relation for clusters and using infrared data from 2MASS (\citealp{2006AJ....131.1163S}) we find a predicted $ L_X = 3.7^{+0.8}_{-0.7}\times 10^{40}\, \si{erg\,s^{-1}}$, which is almost two orders of magnitude lower than measured. Instead the predicted and measured X-ray luminosities are almost consistent when assuming that the galaxy is a field galaxy, which has not lost gas from the interaction with the cluster ICM.

In Figure \ref{fig:smooth} we can also see that a cone at the end of the head (region ``Head'' in Figure \ref{fig:smooth}) of the structure is pointing towards the galaxy. The galaxy has a radio counterpart visible in the NVSS image (\citealp{1998AJ....115.1693C}, Fig. \ref{fig:optical}). Interpreting qualitatively the radio morphology as being due to bent lobes of the radio AGN at the center of CGCG254, the inferred direction of motion (south) is consistent with that inferred from the X-ray morphology at the Comet. Unfortunately the radio signal is very weak (1.7 and 4.6 $\sigma$), so this assertion needs a deeper radio observation to be confirmed. The optical/N-IR source to the south-west of the Comet is a galaxy (J181025.63+495541.7) with a V-band magnitude of $16.6$ (see \citealp{2009A&A...497..667V}) and a redshift of $z=0.051$ (\citealp{2004AJ....128.1558S}), so this galaxy most likely belongs to the cluster but we see no indications of interaction with it or the Comet or CGCG254. 

Compared to the $\SI{70}{kpc}$ X-ray tail of a galaxy analyzed in \cite{2006ApJ...637L..81S}, this X-ray tail is about 20 times more luminous, the Chandra ACIS-I count-rate for our object is 40\% higher than the one of the X-ray tail in \cite{2006ApJ...637L..81S}, while the two objects have a very similar temperature structure. This means the amount of gas that apparently has been stripped of the galaxy, is very high, or the galaxy lost almost all of its gas to the ICM. If this scenario was confirmed by the analysis of a longer observation, one would have the chance to study in detail the properties of this interaction for such a luminous and massive object.

\section{Conclusion}
In our short Chandra observation of Zwicky 8338 we identify a long X-ray tail, of which we study the properties. We measure the surrounding ICM to have a temperature of $\SI{3.6(4)}{keV}$. The X-ray tail itself shows thermal emission with a temperature of $\SI{0.77}{keV}$. Indications point to a head-tail structure within the X-ray tail: A cooler, but more metal rich part has a very peaked brightness distribution, while the other part is more diffuse. The $\SI{14}{mag_V}$ galaxy CGCG254-021 is very close to this X-ray object and several indications point to the scenario that the X-ray tail is stripped gas from this galaxy or from a small galaxy group with this galaxy as its BCG: 
\begin{itemize}
	\item The projected distance between the peak of the X-ray tail and the galaxy center is only $\SI{40}{kpc}$.
	\item The redshift of galaxy and X-ray tail are consistent with the cluster redshift.	
	\item The apparently bent radio lobes from the galaxy's AGN are consistent with the inferred direction of motion.
	\item The galaxy is the brightest galaxy within this region with the highest star formation rate in the past.
	\item The tail's gas mass inferred from the X-ray luminosity is far too low for its estimated total mass, if one compares these values with galaxy group scaling relations.
	\item The stellar mass of the galaxy is consistent with those of BCGs in galaxy groups.
\end{itemize}
With a deeper X-ray observation it would be possible to characterize more detailed properties as well as a definite scenario for the interaction history.
\begin{acknowledgements}
The authors would like to thank Sandra Burkutean, Alberto Doria and Lorenzo Lovisari for helpful discussions. 
GS, THR acknowledge support by the German Research Association (DFG) through grant RE 1462/6. GS acknowledges support by the Bonn-Cologne Graduate School of Physics and Astronomy (BCGS). THR acknowledges support by the DFG through Heisenberg grant RE 1462/5.
\end{acknowledgements}
\bibliographystyle{aa}
\bibliography{Astro}
\end{document}